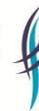

# QUALITY ASSURANCE PRACTICES IN AGILE METHODOLOGY

[1]Almustapha Abdullahi Wakili, [2]Lawan Nasiru Alhasan,
[3]Abubakar Hamisu Kamagata,

[1]Student, MSc. Software Engineering, Department of Computer Application,
Mewar University Chittorgarh, almustaphawakili@gmail.com
[2]Phd Research Scholar, Department of Computer Science Sangam University Bilwarah,
lawannasiruhassan935@gmail.com

[3]Department of Software Computer Science PPSavani University,kamagata012@gmail.com.

**ABSTRACT:**

*The complexity of software is increasing day by day the requirement and need for a verity of software products increases, this necessitates the provision of a strong tool that will make a balance between production and quality. The practice of applying software metrics to the development process and to a software product is a critical task and crucial enough that requires study and discipline and which brings knowledge of the status of the process and/or product of software in regards to the goals to achieve, this discipline is known as quality assurance which is the key factor behind the success of every software engineering project, the quality assurance activities are what result in the qualitative product as well as the process in both conventional software development methodology and agile methodology. However, agile methodology is now becoming one of the dominant method adopted by most of the software industries because it allows developing of software with very limited requirement and supports rapid changes in the requirement, the method may produce the product very fast but we might not guarantee the quality of the product unless we apply the SQA activities to the process. This research paper aimed to study the quality assurance activities practice in agile software development methodology, investigate the common problems and key drivers of quality in agile, and propose a solution to improve the practice of SQA in agile methodology by analyzing the parameters that assure quality in agile software.*

## 1.0. INTRODUCTION

The main purpose of software engineering is to provide a process and principles for building a qualitative software that is delivered on time, within budget and compile with its requirement, this software engineering principle ease the process of building the software product and reduce the rate of software product failure, however, building a qualitative software require more critical strategies in which failure to adopt them carefully still leads to a verity of software product failure. However, building a qualitative software requires more critical strategies in which failure to adopt them carefully still leads to a verity of software product failure. The Varsity and complexity of software increased from day by day, which necessitate the provision of technique that will make a balance between quality and productivity. The practice of correlating software metrics to a software product and to a software process is a complex task that requires study and discipline and which brings knowledge of the status of the process and/or product of software in regards to the goals to achieve, this discipline is known as quality assurance and management which are the key factors that contribute to the success of every software engineering project, the quality assurance activities are what result in a qualitative product as well as the process in both conventional software development methodology and agile methodology

Companies spend billions of dollars on IT projects each year, with average spending at 4 to 5 percent of their annual revenue on IT (Charette, 2005). In 2005, companies and organizations spent an estimated $1 trillion on Information Technology hardware, software, and services worldwide (Charette, 2005). Project failures have cost the US economy at least $25 billion and maybe as much as $75 billion (Charette, 2005). Based on the recent project failure statistics, companies have wasted billions of dollars on IT projects alone. Many researchers have written about the causes of software project failure and have come up with suggestions and critical success factors for effective management of projects. This research takes a different approach by examining how software quality assurance practices in agile methodology can impact project outcomes. It is author's contention that software quality assurance (SQA) plays a vital and critical role in the life cycle of building a software, known as the software development life cycle (SDLC) and can impact the overall success of a project. Failure to focus to SQA can result in budget overruns, schedule delays, failure to meet project objectives and requirements, and poor customer satisfaction (Chow, 1985). In fact, a quality is considered as the most important and vital requirement of software products, a business essential, a competitive necessity, and even a survival issue for the soft-ware industry (Murugesan, 1994).

## 1.1 BACKGROUND OF THE STUDY

Notwithstanding that, from what some scholars have found, we have seen how Software Quality Assurance play a very vital and critical role in the success of every software products, for us to study the practice of quality assurance in the agile methodology we need to clearly understand the concept of software quality and quality assurance and we indeed require the knowledge of Software development life cycle SDLC and agile development process. In software development industries the concept of quality is defined and expressed in many different ways, some scholars definitions focus on error-free functionality of software product whereas some scholars are found to emphasize on customer satisfaction with the developed software product, but it is really hard to cite an absolute definition of quality even after going through literature. Verily it is never easy defining "quality" you can simply take it as it perfectly suite serving your purpose, nevertheless it plays a very vital role when it comes to meeting the expectation of your client, so considering the quality of a product or service is a key to the success of your team, company or an organization.

Quality management is a process of ensuring that the required level of quality is achieved not only in products but also in quality in the process through which these products are produced. It involves defining some appropriate quality standards and procedures to ensure that these are followed. The aim should be to develop a 'quality culture' where achieving quality is seen as the responsibility of everyone in the organization. Quality planning, quality assurance, and quality control are the key activities which are involved in quality management. (W.E.Lewis)

Current findings seem to express that Plan-oriented projects or the projects following the conventional development process, are influenced under SQA activities. Agile software development methods have changed the scenario of conventional software development. Manifesto for Agile Software Development (NASA, 2020)emphasizes close team collaboration, quality, and the relationship between developer and customer. According to Miller, one of the characteristics of the agile software process is "People-oriented, i.e. agile processes favor people over processes and technology" So, if agile development methodology is more people-centric, then the effective role of SQA in projects that adopt agile development; needs to be more comprehended.

Sajid Ibrahim Hashmi and JongmoonBaik carried out a comparison between XP and the spiral model and the focus was on quality assurance. They claim that in agile development, developers may also be responsible for QA activities(Sajid S. I. and Jongmoon B,). Agile development projects consist of short iterative development and release of products. And projects, following agile development, evolve around the developer and customer who are responsible to maintain product quality (E. Mnkandla, .B. Dwolatzky).If responsibility for the quality, in agile development, is shifted on customer and developer, then the supporting role of QA must be identified. SQA is not only res possible for a particular project but also maintains the processes and culture of the organization. In (Charles Field et al,, 2007)User Experience Design (UXD) team approach is introduced collaborating with developers, this UXD team approach seems an attempt to redefine and replace the role of SQA in agile projects. The main focus of this paperwork is to highlight some gaps in agile SQA activities and to put forward suggestions for improvement by answering the following questions.

1) What are the key drivers to quality assurance in agile methodology?
2) How can we manage and control the quality of the agile process?
3) What QA practice in agile will ensure quality software

**1.2 RESEARCH GOAL**

Thus, as we have seen earlier there is need to propose solutions to the problems in containing quality assurance in agile software development, and there is need to feel the gaps, the main goal of the research paper is to answer the above-asked questions and analyze the quality assurance practices' in agile methodologies, their abilities, and their frequency

**2.0. LITERATURE REVIEW**

Agile processes of software development, such as Scrum, extreme programming (XP), etc., rely on best practices that are considered to improve quality in software development. It can be said that best practices aim to induce software quality assurance (SQA) into the project. The quality assurance (QA) activities, in the software development process, are also the backbone of the project. (Sagheer Maria, Zafar Tehreem, Sirshar Mehreen, 2015)

Agile methodology is presented since the '90s in many books, articles journals, etc. but a few kinds of research are conducted upon the quality assurance in agile software development methodology. The main purpose of this paper is to search for the answer that how quality is assured in agile software development. CMMI ensures many standards regarding the modification of agile methods but the main reason for introducing CMMI was to provide a standard that will be suitable for all modern iterative methodologies. Using qualitative and quantitative research method data is collected from agile practices about the common practices followed in agile development. The main advantage of this is that it provides a successful quality assurance model for the agile projects. (Noura, G. M. Andrew and W. B. Gray, 2007)

Due to the fact that software development is changing in nature, quality professionals must change with it. Quality is the basic aspect of agile methodology, which is tested by the developers and the customer to have a better quality of the system. This technique will improve the quality but lessen the participation of the quality assurance team. A true agile framework is best than the traditional one because the testing and error fixing is much easier and quicker. Using this approach all the testing is done on the developer's end but acceptance and usability testing can be done on users' end. Despite all the advantages, achieving a true agile quality assurance flow is not easy and it requires coordination among stakeholders.

Agile focuses on accelerated and less costly software development. Achieving both this technique put somewhat compromise in the quality and will unable to provide the reusability of its software developed parts. In computer engineering as well as in software engineering reusability is the important factor of the source code, which is then used to add more functions to that system having no or some modification. With reusability, the productivity of the developers is increased as well as with the increase in reliability and the maintainability of the software. Through the following threw ways the reusability is added in the agile development that are: Component-based development, Reusable designs, and refactoring the design patterns. According to the proposed model, searching will become faster in agile enhancing reusability. Pattern-based designs, UML designing, and analysis are incorporated. Agile development includes quality factor but is unable to provide reusability of its modules. ( S. Sukhpal and C. Inderveer, 2012)

An Assessment of Perception of Values of Professionals on Quality (S. Mariana and H. Paulo, 2011). In the current scenario, the agile technique becomes very important. This paper presents an analysis of the survey report which shows the relationship between the use of agile practices and the quality of software products. This study suggests that the practitioners should use a combination of agile practices and it will improve the understanding of software development. This research suggests the professional's perception and the working environment and doing the work in the defined time. Delivering high-quality software that meets its requirement in time or in the defined deadline becomes a challenge and this suggests that the organized set of agile practices should be adopted to achieve a better-quality product. ( A.S. Nookabadi and J.E. Middle., 2010)

The agile technique produces software faster and enhances quality. The paper focuses on the quality factors and depicts how these factors enhance the quality of software. An agile software life cycle is a series of steps that show the software quality process. Response to the variable requirements, customer satisfaction level, and the continuous delivery of the software is the major advantage of agile while its disadvantage is that it is difficult to access the effort (in terms of time and cost) required at the beginning of the life cycle. In summary agile methodologies enhances the flexibility of the software system. (H. Amran, M. A. Kashem and S. Sahelee, 2013)

With a large paradigm shift in the software industries, a number of distinct software development methodologies have been proposed. Along with these methods, software quality methods and techniques have also been evolved. In this research paper, q quality matrix for the agile development has been shown which will ensure the quality of the product being developed. In this matrix, eight quality attributes along their attributes have been mentioned which depict the role of these attributes in all the phases of SDLC. Analyzing the matrix, it has been noticed that the most important quality attribute is flexibility while portability and understandability comes afterward. (M. Usman, M. Haseeb and J. Ali , 2014)

## 3.0. EVALUATION OF QUALITY IN AGILE PROCESS

However, we can evaluate quality assurance practices in agile processes. which can be done through:

- The provision of clear and detailed knowledge about specific quality issues in the agile processes.
- Identifying innovative ways to improve quality in agile.
- Identifying specific agile quality techniques for specific agile methodologies.

| Technique | Description |
|---|---|
| Refactoring | Make small changes to code, Code behavior must not be affected, Resulting code is of higher quality (Ambler, 2005) |
| Test-driven development | Make small changes to code, Code behavior must not be affected, Resulting code is of higher quality (Ambler, 2005). |
| Acceptance testing | Create a test, Run the test, Make changes until the test passes (Ambler, 2005). |
| Continuous integration | Quality assurance test done on a finished system, Usually involves the users, sponsors, customer, etc. (Huo, Verner, Zhu, & Babar, 2004). |
| Pair programming | Done on a daily basis after developing a number of user stories. Implemented requirements are integrated and tested to verify them. |
| | This is an important quality feature. Two developers work together in turns on one PC, Bugs are identified as they occur, Hence the product is of a higher quality (Huo et al., 2004). |
| Face-to-face communication | Preferred way of exchanging information about a project as opposed to use of telephone, email, etc. Implemented in the form of daily stand-up meetings of not more than twenty minutes (Huo et al, 2004). This is similar to the daily Scrum in the Scrum method. It brings accountability to the work in progress, which vital for quality assurance. |
| On-site customer | A customer who is a member of the development team, Responsible for clarifying requirements (Huo et al., 2004). |
| Frequent customer feedback | Each time there is a release the customer gives feedback on the system, and result is to improve the system to be more relevant to needs of the customer (Huo et al., 2004). Quality is in fact meeting customer requirements. |
| System metaphor | Simple story of how the system works (Huo et al., 2004), Simplifies the discussion about the system between customer/ stakeholder/ user and the developer into a non-technical format. Simplicity is a key to quality |

Table 1, agile quality technique as applied in an extreme programing

Literature shows that Huo et al. (2004) developed a comparison technique whose aim was to provide a comparative analysis between quality assurance in the waterfall software development model (as a representative of the traditional camp) and quality in the agile methodologies. The results of the analysis showed that there is indeed quality assurance in agile development, but it is achieved in a different way from the traditional processes. The limitations of Huo et al.'s tool, nevertheless, are that the analysis:

• Singles out two main aspects of quality management namely quality assurance and verification and validation.

• Overlooks other precious QA techniques used in agile processes to achieve higher quality management.

• Agile quality assurance requires a step beyond the traditional software quality assurance approaches.

Another challenge of Huo et al.'s technique is that while the main purpose of that analysis was to show that there is quality assurance in agile processes, it does not make it clear what the way forward is. Agile proponents do not seem to be worried about the comparison between agile and traditional processes as some of the more zealous "agilists" believe that there is no way traditional methods can match agile methods in any situation (Tom Poppendieck, personal e-mail 2005). So, the evolution described here improves on (Huo et al., 2004).

| Software Quality Parameters | Agile Techniques | Possible Improvements |
|---|---|---|
| Correctness | Write code from minimal requirements. Specification is obtained by direct communication with the customer. Customer is allowed to change specification. Test-driven development. | Consider the possibility of using formal specification in agile development, Possible use of general scenarios to define requirements (note that some development teams are already using this). |
| Robustness | Not directly addressed in agile development. | Include possible extreme conditions in requirements. |
| Extendibility | A general feature of all OO developed applications. Emphasis is on technical excellence and good design. Emphasis also on achieving best architecture. | Use of modeling techniques for software architecture. |
| Reusability | A general feature of all OO developed applications. There are some arguments against reusability of agile products (Turk, France, &Rumpe, 2002; Weisert, 2002). | Develop patterns for agile applications. |
| Compatibility | A general feature of all OO developed applications. | Can extra features be added for the sake of compatibility even if they may not be needed? This could contradict the principle of simplicity. |
| Efficiency | Apply good coding standards. | Encourage designs based on the most efficient algorithms. |
| Portability | Practice of continuous integration in extreme programming | Some agile methods do not directly address issues of product deployment. Solving this could be to the advantage of agility. |
| Timeliness | Strongest point of agility, Short cycles, quick delivery, etc. | |
| Integrity | Not directly addressed in agile development. | |
| Verifiability | Test-driven development is another strength of agility. | |
| Ease of use | Since the customer is part of the team, and customers give feedback frequently, they will most likely recommend a system that is easy to use. | Design for the least qualified user in the organization. |

Table 2. Mapping Software Quality Parameters to agile technique

**AGILESOFTWARE PROCESS IMPROVEMENT**

A clear view of agile processes leads to a notion that agile methodologies are a group of processes and activities that reduces the timeframe for developing a software program and introduced innovative techniques for embracing rapidly changing business requirements. With time, these relatively new techniques of QA should develop into mature software engineering standards. However, we have seen that Enabling Reusability in Agile Software Development (C. Inderveer et al, 2012) 2.8) Agile Practices: 2.9) is what makes the process of building software more time effective.

**5.0. THE AGILE METHODOLOGY EVALUATION FRAMEWORK**

All agile methodologies have striking similarities amongst their processes because they are based on the four agile values and 12 principles. It is interesting to note that even the authors of agile methodologies no longer emphasize their methodology boundaries and would use practices from other agile methodologies as long they suit a given situation ( (Beck, K. et al., 2004). In fact, Kent Beck in his extreme programming (XP) masterclasses frequently mentions the errors of extremism in the first edition of his book on XP (Beck, 1999). A detailed review of agile methodologies reveals that agile processes address the same issues using different real-life models. The evaluation technique presented in this paper reveals, for example, that lean development (LD) views software development using a manufacturing and product development metaphor. Scrum views software development processes using a control engineering metaphor. Extreme programming views software development activities as a social activity where developers sit together. Adaptive systems development (ASD) views software development projects from the perspective of the theory of complex self-adaptive systems (Mnkandla, 2006). Tables 1 to 2 summarize the analysis of agile methodologies. Only a few of the existing agile methodologies have been selected to illustrate the evaluation technique. There is a lot of subjectivity surrounding the choice of methodology elements. It is not within the scope of this paper to present a complete taxonomy of methodologies. For more detailed taxonomies see Avison and Fitzgerald (2003), Boehm et al. (2004), Glass and Vessey (1995), and Mnkandla (2006). Therefore, the elements used here were chosen to reveal the similarities amongst different agile methodologies. The importance of revealing these similarities is to arm the developers caught up in the agile methodology jungle wondering which methodology to choose. While the methodology used in your software development project may not directly lead to the success of a project and may not result in the production of a high-quality product use of a wrong methodology will lead to project failure. Hence, there is in wisdom

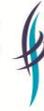

## 6.0. CONCLUSIONS

Even though some agile practices are not new, agile methods are recent and have become very popular in the industry. There is an important need for developers to know more about the quality of the software produced. Developers also need to know how to revise or tailor their agile methods in order to attain the highest level of quality they require. In this paper, we have analyzed quality assurance practices' in agile methodologies, their abilities, and their frequency. The conclusion we draw here is: 1) most of the agile methods practices have quality assurance abilities, some of them are inside the development phase and some others can be separated out as supporting practices 2) the frequency with which this agile Quality Assurance practices occur is higher than in a waterfall development 3) agile QA practices are supported in very early phases of agile due to the agile process characteristics. From this analysis, we identified some issues for which development criteria might be desirable. According to the process quality, a team requires and time they have available they can tailor agile practices. However, is difficult, sometimes even not realistic to compare the software quality resulting from the use of a waterfall model with agile methods because their initial development conditions, especially the cost, are not comparable